%% file: main.tex
\documentclass[%
reprint,
superscriptaddress,
 feyn,feynmf,
 amsthm,amsmath,amssymb,
 aps,
prd,
longbibliography,
]{revtex4-1}
\usepackage[utf8]{inputenc}

\usepackage{xcolor} 

\usepackage{mathrsfs}
\usepackage{subcaption}
\usepackage{adjustbox}
\usepackage{graphicx}
\usepackage{dcolumn}
\usepackage{bm}

\usepackage[T1]{fontenc}
\usepackage{booktabs, array, mathptmx, float, tabularx, booktabs, lipsum, amsmath,multirow}
\usepackage{siunitx, xcolor}
\usepackage[version=4]{mhchem}
\usepackage[colorlinks,linkcolor=blue,anchorcolor=blue,citecolor=blue]{hyperref}

\begin{document}
\title{A search for two-component Majorana dark matter in a simplified model using the full exposure data of PandaX-II experiment}

\input{authorlist}


\begin{abstract}
In the two-component Majorana dark matter model, one dark matter particle can scatter off the target nuclei, and turn into a slightly heavier component. In the framework of a simplified model with a vector boson mediator, both the tree-level and loop-level processes contribute to the signal in direct detection experiment.
In this paper, we report the search results for such dark matter from PandaX-II experiment, using total data of the full 100.7 tonne$\cdot$day exposure. 
No significant excess is observed, so strong constraints on the combined parameter space of mediator mass and dark matter mass are derived. With the complementary search results from collider experiments, a large range of parameter space can be excluded.
\end{abstract}

\maketitle

\section{Introduction}
\label{sec:introduction}
Dark matter (DM) can be used to explain the missing mass in our universe~\cite{bertone2005particle}, thus understanding it becomes one of the fundamental questions of modern physics.
Though the nature of DM remains unknown, treating it as a set of particles beyond the standard model (BSM) of particle physics is one of the most discussed scenarios. 
Among numerous theoretical models, the hypothesis of weakly interacting massive particles~(WIMPs), existing in many BSM theories, is very promising~\cite{Arcadi:2017kky}.
WIMPs could be detected directly through their elastic scattering with nucleus~\cite{Goodman:1984dc,Schumann2019}.
The xenon based experiments, such as PandaX-II~\cite{PandaX:2016pdl,PandaX-II:2020oim,Zhao:2020ezy}, LUX~\cite{LUX:2013afz},  XENON1T~\cite{XENON:2015gkh1} and Pandax-4T~\cite{PandaX:2018wtu,PandaX-4T:2021bab}, have given the most stringent constraints on the spin independent~(SI) WIMP-nucleon elastic scattering for a wide range of WIMP masses.
However, the null results motivate the consideration of other possible interaction mechanism between DMs and nucleons.  
There may be natural methods to reduce the elastic scattering rate while leaving the annihilation rate sufficiently large to avoid overclosing the Universe. One example is the two-component Majorana DM model with very small Majorana mass splitting relative to the Dirac mass~(also known as the pseudo-Dirac DM or pDDM), where the lighter component is stable and represents the DM candidate~\cite{DeSimone:2010tf,Davoli:2017swj,Bramante2016,Chao:2020yro}. 

We consider a simplified model as a concrete realization of the pDDM scenario, where a new gauge boson mediator $V$ is introduced to connect the two components with the standard model particles. This scenario gives
interesting phenomena in both the direct detection and collider experiments~\cite{Bell2018}. 
In the direct detection experiments, the light component as the DM candidate interacts with the target and becomes the heavy component at tree level. 
The minimum kinetic energy requirement on the incoming particle to produce a heavier final particle leads to a distinct nuclear recoil~(NR) energy spectrum offset from zero, a signature which is more generally referred to as the inelastic DM scattering, or iDM~\cite{Tucker-Smith:2001myb,Tucker-Smith:2004mxa,Bramante2016,cui2009candidates}. There have been a number of searches performed on this inelastic scattering~\cite{XENON10:2009sho,Chen2017}.
However, with this mediator $V$, elastic scattering can also happen, as pointed in Ref.~\cite{Bell2018}, but only at loop level, which could become the dominating process when the tree level process is kinematically suppressed. 
In this paper, we perform a search for pDDM through both the inelastic and elastic processes using the full exposure data of PandaX-II, and provide constraints on the simplified vector-mediator mass parameter space.

\begin{figure*}[htbp]
    \centering
    \begin{subfigure}[t]{.2\textwidth}
    \includegraphics[width=1.0\textwidth]{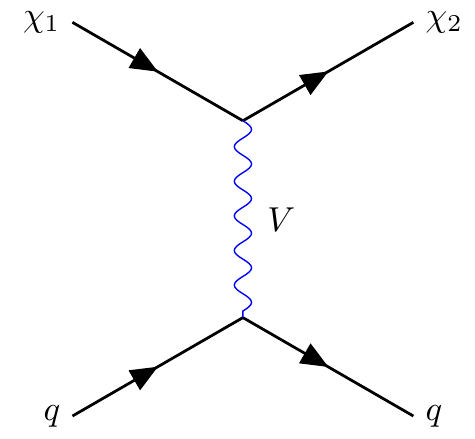}
    \caption{Tree-level}
    \label{fig:idm_tree}
    \end{subfigure}
    \begin{subfigure}[t]{.7\textwidth}
    \includegraphics[width=0.4\textwidth]{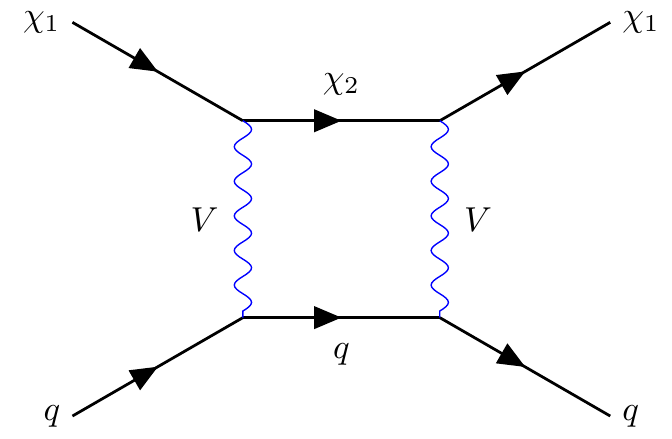}
    \includegraphics[width=0.4\textwidth]{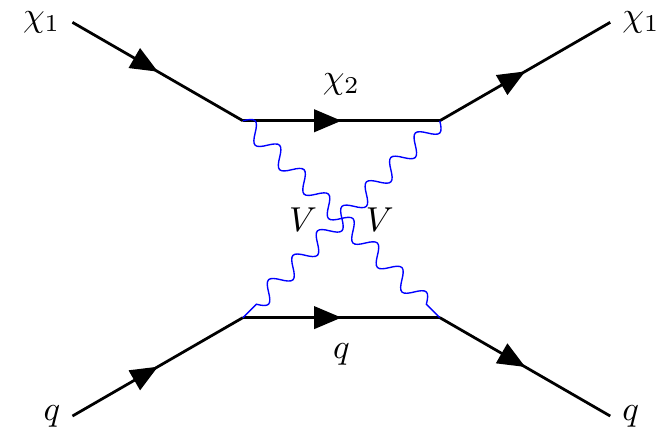}
    \caption{Loop-level}
    \label{fig:idm_loop}
    \end{subfigure}
    \caption{Interaction between DM and quarks in upscattering pDDM model. (a) Tree level process and the mass splitting $\delta$ is limited in direct detection search; (b) Loop level processes, generating spin independent elastic scattering.~\cite{Bell2018}}
    \label{fig:fm}
\end{figure*}

\section{Signals in xenon detector}
\label{sec:model}
In the simplified model for the pDDM scenario, the tree and loop level scattering processes are illustrated in Fig.~\ref{fig:fm}. The tree-level diagram can be described by $t$-channel exchange of the vector boson $V$,
and the loop diagrams are generated by double exchange of $V$.
In the leading order tree-level process, the DM particle $\chi_{1}$, scattering off the target nucleus, can be transformed into a slightly heavier mass state $\chi_{2}$ with a mass splitting $\delta=m_{\chi_2}-m_{\chi_1}$, and therefore inelastic scattering emerges. The next-to-leading order loop-level processes give elastic scattering instead.
The fractions of the different levels processes vary with several independent parameters $\left\{m_{\chi_1}, \delta, v\right\}$, where $v$ is the incoming DM velocity.

For the processes to the pDDM scattering phenomenon, in the direct detection with relatively small momentum, the effective Lagrangians are given as below~\cite{Bell2018}
\begin{align} 
  \label{eq:L_tree}
L_{\rm tree} &= \frac{g^{2}}{M^{2}}\bar{\chi}_{1}\gamma^{\mu}\chi_{2}\bar{q}\gamma_{\mu}q \rightarrow c^{\rm N}_{\rm 5}\bar{\chi}_{1}\gamma^{\mu}\chi_{2}\bar{N}\gamma_{\mu}N,\\
\label{eq:L_loop}
L_{\rm loop} &= \frac{4g^{4} m_{\chi_1}m_{\rm q}}{16 \pi^{2} M^{4}} F_{3}(\frac{m_{\chi_1}^{2}}{M^{2}})\bar{\chi}_{1}\chi_{1}\bar{q}q \rightarrow c^{\rm N}_{\rm 1}\bar{\chi}_{1}\chi_{1}\bar{N}N,
\end{align}
where $g^2$ is the product of the coupling constant of the mediator to DMs~($g_{\chi}$) and the flavour-universal coupling to quarks~($g_q$), 
$M$ is the mass of the mediator, and the function $F_3$ takes the form of 
\begin{equation}
    F_{3}(x) = \frac{(8x^{2}-4x+2)\log(\frac{\sqrt{1-4x}+1}{2\sqrt{x}})+\sqrt{1-4x}[2x+\log(x)]}{4\sqrt{1-4x}x^{2}}.
\end{equation}
The quark-level effective Lagrangians can be matched to the nucleon-level ones with couplings $c_i^N$ calculated using the Mathematica package, DirectDM~\cite{Bishara2017}. 
DirectDM takes the Wilson coefficients of the relativistic effective theory describing the interaction of DM with quarks as inputs, and matches them into an effective theory of DM-nucleon interaction. The spin-independent DM-nucleon cross section $\sigma_{n}$ depends on the effective couplings~\cite{Li2019_1},
\begin{equation}
\sigma_{n}=\frac{(c^{N}_i\mu_{n})^{2}}{\pi},
\label{eq:sigma}
\end{equation}
where $\mu_{n}$ is reduced mass of the DM and the nucleon, and the index $i$ takes the value of 5 and 1 for the tree and loop level processes, respectively.

\begin{figure}[htbp]
  \centering
  \begin{subfigure}{0.4\textwidth}
  \includegraphics[width=0.9\textwidth]{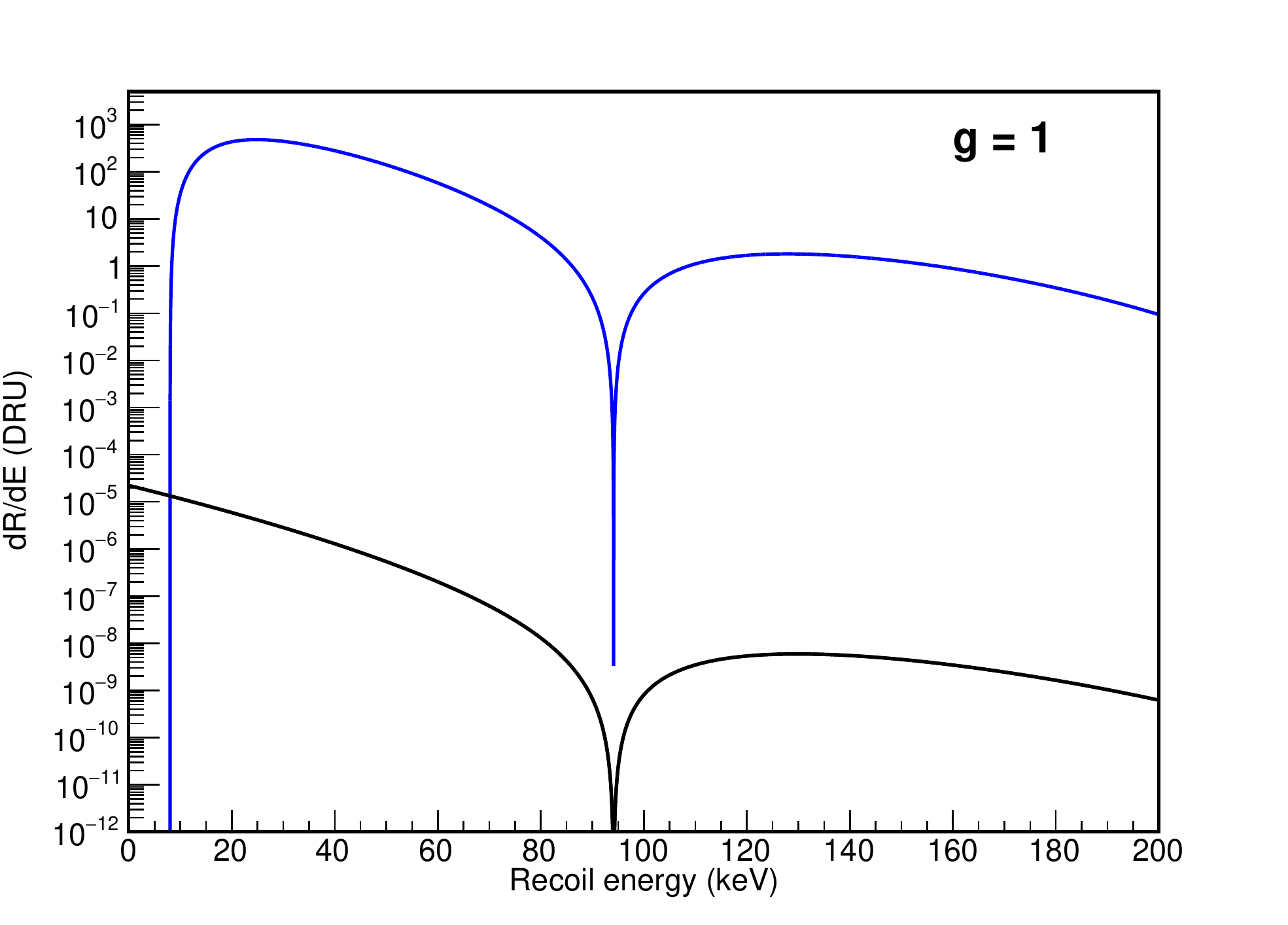}
  \caption{$m_{\chi}$=100~GeV}
  \label{fig:100}
  \end{subfigure}
  \begin{subfigure}{0.4\textwidth}
  \includegraphics[width=0.9\textwidth]{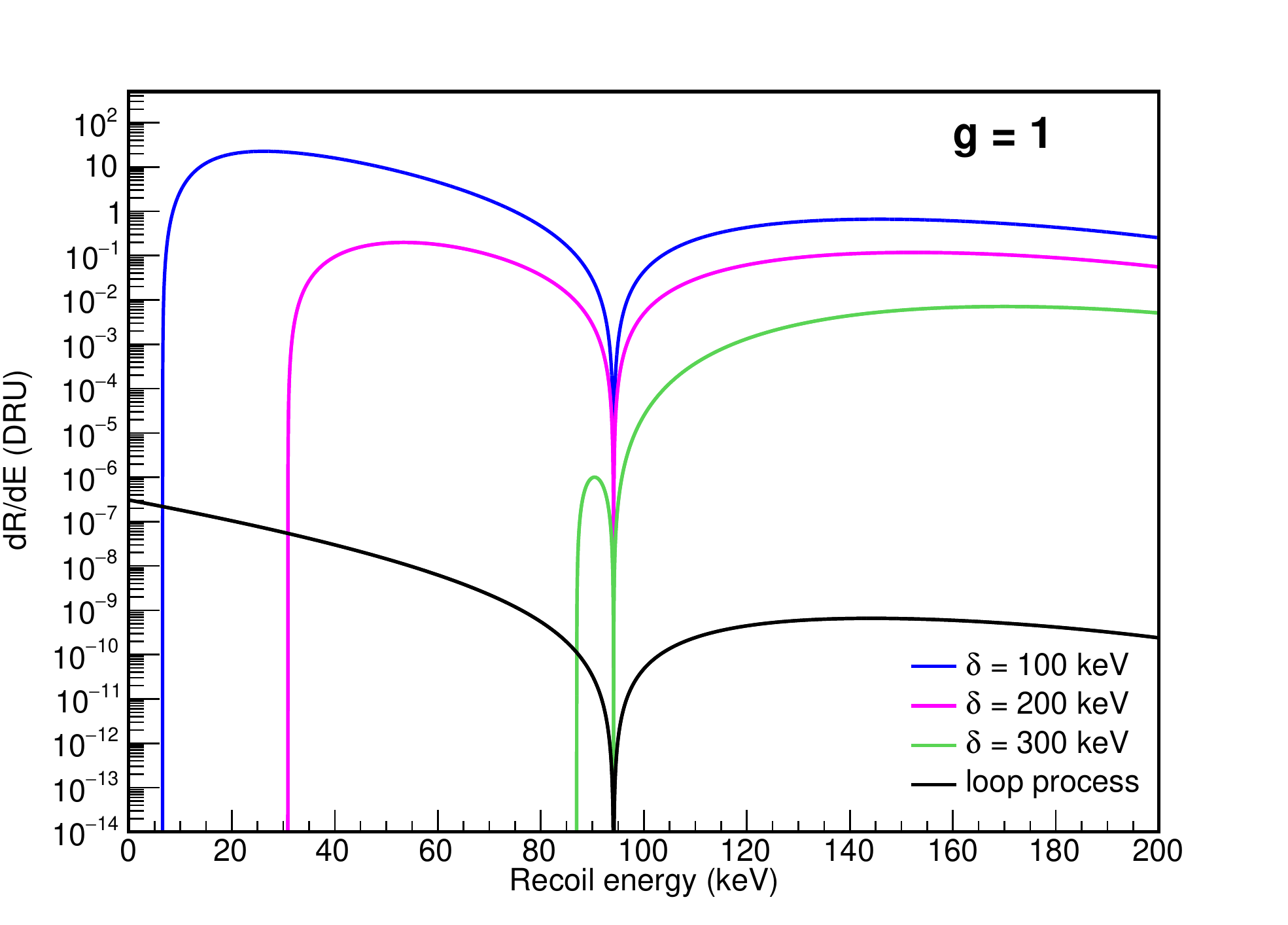}
  \caption{$m_{\chi}$=10~TeV}
  \label{fig:10000}
  \end{subfigure}
  \caption{The expected event rate for DM mass of 100~GeV (\subref{fig:100}) and 10~TeV~(\subref{fig:10000}), assuming $M = 150~$GeV, and $g= 1$. For the tree level process, the rates calculated with three different mass splitting $\delta$ of 100 (blue), 200 (magenta) and 300 (green)~keV are plotted. The loop level rate is shown as black line. For the DM mass of 100~GeV, the tree level processes of $\delta=200, 300$ keV are absent due to kinematic limit from the halo escape velocity. The astrophysical parameters in Ref.~\cite{Baxter:2021pqo} are used in this calculation.}
  \label{fig:sig} 
\end{figure}

In the tree-level process, the incoming DM needs a certain kinetic energy to produce the heavier component after scattering. Therefore this inelastic scattering cross section has a kinematic suppression depending on the mass splitting $\delta$. 
The loop-level cross section is generally suppressed in comparison with the tree level, due to the dependence on higher order of the mediator mass $M$. However, the kinematic suppression does not exist for the loop-level elastic scattering process.
Therefore, for small DM masses with a large mass splitting, when the tree level process is significantly suppressed, the loop level process becomes dominant in the low recoil energy region. This behavior is illustrated in Fig.~\ref{fig:sig}, where the expected event rates of the tree-level and loop-level process for DM masses of 100~GeV and 10~TeV are presented.

The detailed formula of signal rate of tree-level inelastic scattering is discussed in Ref.~\cite{Chen2017}. We describe the key points of the inelastic process here.
Due to the mass difference $\delta$ between the outgoing and the incoming DM, the velocity of incoming DM needs to exceed a threshold,
the global minimal velocity $v_{\text{min}}^{\text{global}}$,  to allow such an interaction.
\begin{equation}
  \label{eq:v_min_eq}
  v_{\text{min}}^{\text{global}} = \sqrt{2\delta/\mu},
\end{equation}
where $\mu$ is the reduced mass of the DM and the target nucleus. However, for a certain mass splitting, the $v_{\text{min}}^{\text{global}}$ of the light DM may exceed the galactic escape velocity, where the interaction rate drops to zero. For a detectable recoil energy $E_R$, the velocity threshold is further raised to 
\begin{equation}
v_{\text{min}}=\frac{1}{\sqrt{2E_{\text{R}}m_{\text{T}}}} (\frac{E_{\text{R}}m_{\text{T}}}{\mu}+\delta),
\label{eq:vmin}
\end{equation}
where $m_\text{T}$ is the mass of target nucleus. 
In addition, the usual DM signal selection window in direct detection experiments are within approximately 100 keV NR energy. Therefore, we focus on the mass splitting $\delta$ up to 300~keV, beyond which either the interaction rate diminishes due to halo escape velocity or the recoil energy is out of the signal selection window. 

The differential scattering rate can be parameterized as
 \begin{equation}
 \frac{dR}{dE_{\text{R}}}=\sigma_{n}\frac{A^2\rho_{\chi}}{2m_{\chi}\mu^2}F^2(E_R)\eta(E_R), 
 \label{eq:dRde}
 \end{equation}
 where A is the nucleon number of target element, ${\rho_{\chi}}$ denotes the local DM number density and 
 $\eta(E_R)$ describes the mean inverse velocity of the incoming DM.
For a given $E_R$, $F(E_R)$ is the nucleus form factor, and the Helm form factor~\cite{Lewin:1995rx} is used in this work.

\section{Data selection and backgrounds}
The PandaX-II experiment operates a dual-phase time projection chamber~(TPC) with 580~kg liquid xenon in the sensitive volume. Two arrays of photomultipliers located at the top and the bottom are used to detect the scintillations~(denoted in the unit of PhotoElectron, PE) inside the detector. A detailed description of the detector of the PandaX-II experiment can be found in Ref.~\cite{Tan2016}.
The experiment started running from the late 2015, and ceased operation in the middle of 2019, with about 400 live days of data accumulated~\cite{PandaX-II:2020oim}.

\begin{figure}[htbp]
    \includegraphics[width=0.9\linewidth]{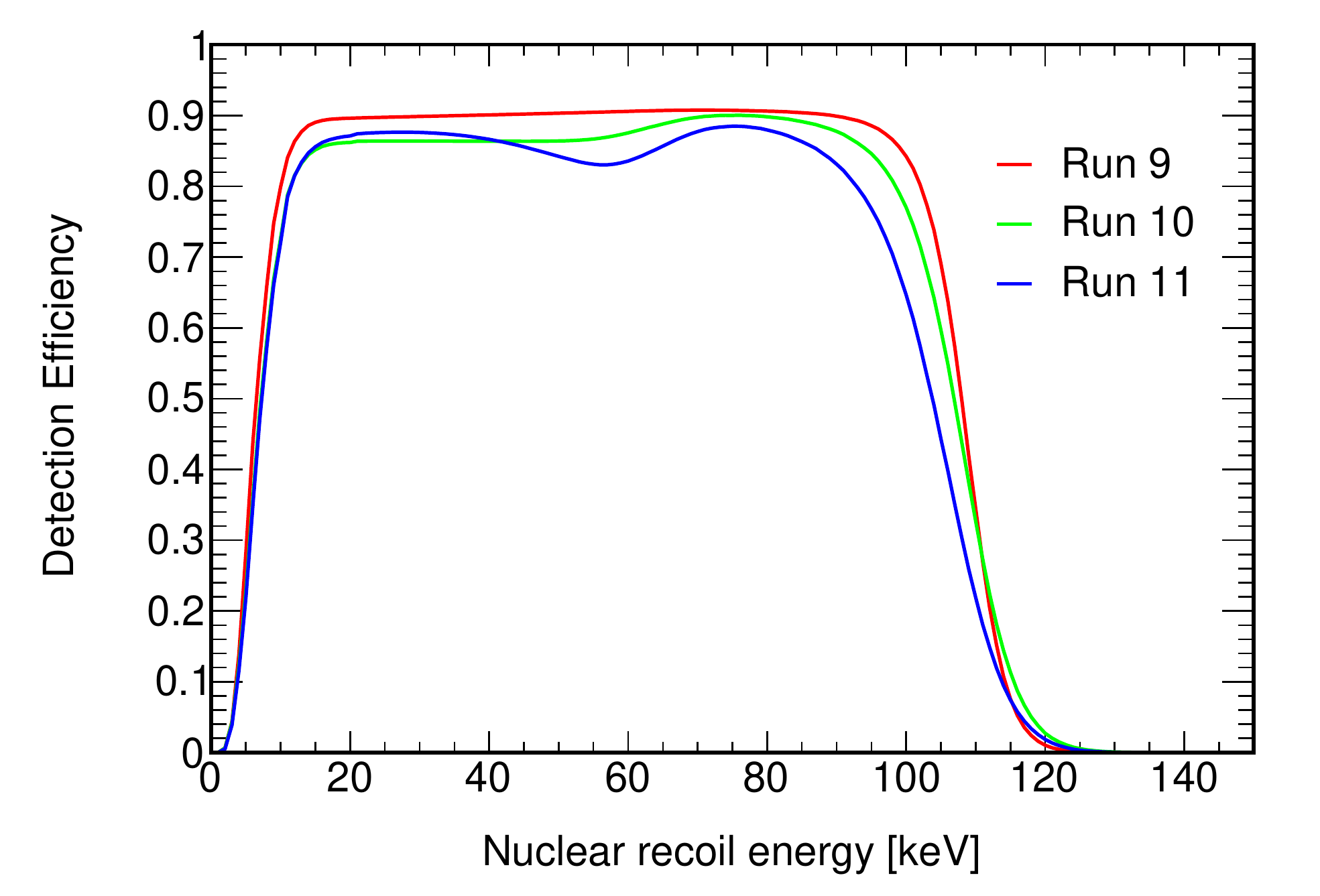}
\caption{PandaX-II detection efficiencies for single scattering NR events as functions of the NR energy.
Run sets are marked with different colors, Run9~(red), Run10~(green) and Run11~(blue).}
\label{fig:NReff}
\end{figure}

The data analysis is performed based on the run 9, 10 and 11 data sets of PandaX-II, with the same data quality cuts and signal correction procedures in Ref.~\cite{PandaX-II:2020oim} applied. 
For each single scattering event with a pair of $S1$~(prompt scintillation light) and $S2$~(ionized electrons) signals, the electron recoil (ER) equivalent energy~(in the unit of keV$_{\rm ee}$) is reconstructed using the same parameters for photon detection efficiency, electron extraction efficiency and single electron gain in Ref.~\cite{PandaX-II:2020oim}. 
The data selection requires the reconstructed energy $E_{\text{rec}}<25$~keV$_{\rm ee}$, $S1>3$~PE and the $S2_{\text{raw}} > 100$~PE.
The signal window is larger than that used in Ref.~\cite{Chen2017} which ensures higher signal acceptance up to 100~keV NR energy. All the quality cuts are applied for the events in the extended signal region except the boosted decision tree~(BDT) cut, which is introduced to suppress the accidentally paired events~\cite{PandaX-II:2022waa}. Since there is negligible accidental background for $S1>100$~PE, the BDT cut is applied for $S1$ below 100~PE only.
With the signal model introduced in Ref.~\cite{PandaX-II:2021jmq}, the overall event detection efficiencies as functions of the NR energy for the three runs are presented in Fig.~\ref{fig:NReff}. The valley of the efficiency curve around 60~keV, especially for run 11, is due to the BDT cut mentioned above.

\begin{figure}[htbp]
\centering
\includegraphics[width=0.9\linewidth]{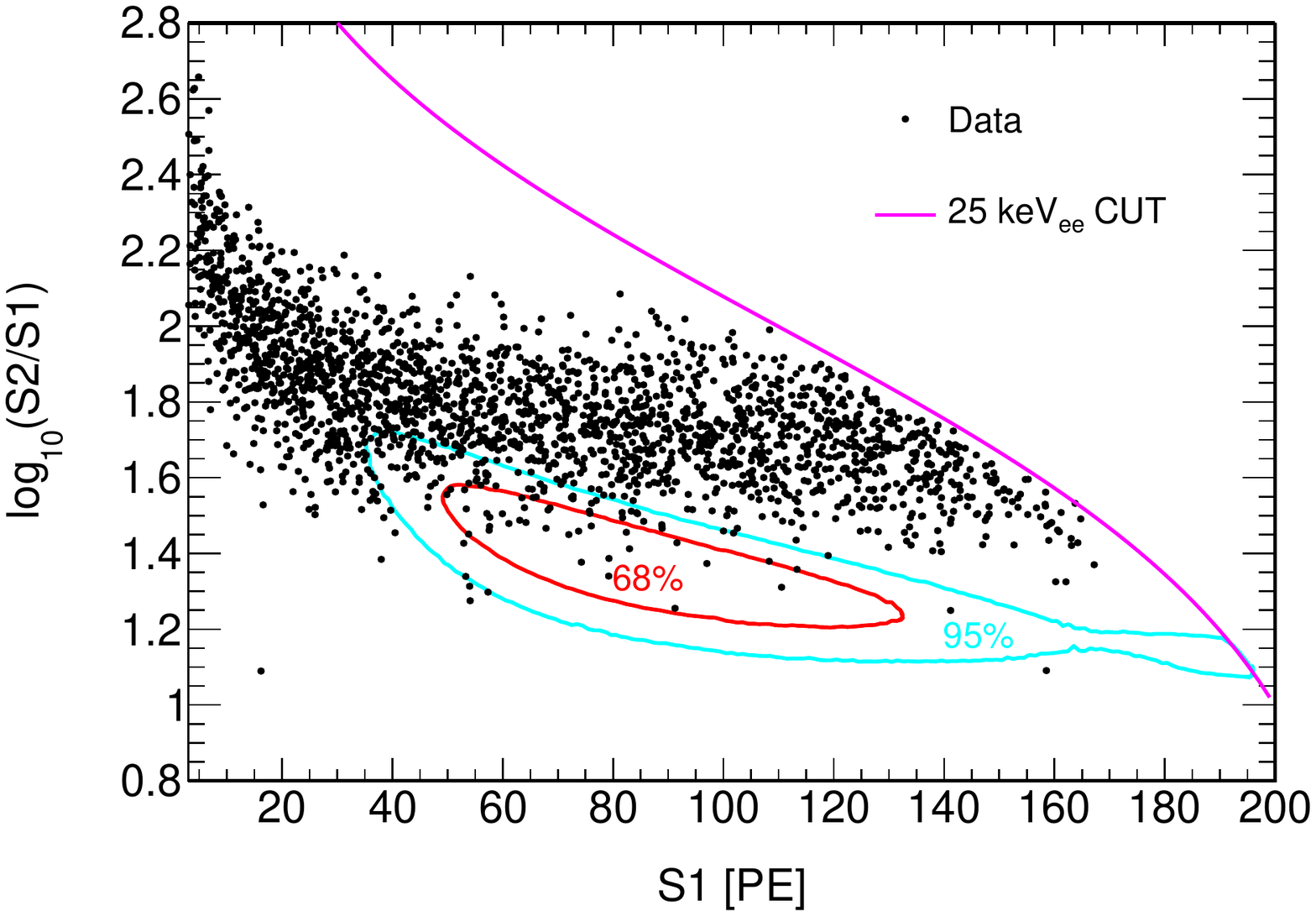}
\caption{$S1$ vs. $\log_{10}(S2/S1)$ distribution of selected events, on top of the expected signal distribution of inelastic scattering with $m_\chi=1$~TeV and $\delta=200$ keV. The events are plotted as black dots.
The red$\slash$cyan curve shows the 68$\%$ $\slash$ 95$\%$ contour of signal distribution.
The magenta line indicates the 25~keV$_{\rm ee}$ upper cut on the reconstructed energy in Run 10.}
\label{fig:2D}
\end{figure}

The fiducial volume~(FV) selection in Ref.~\cite{Zhou2021} is inherited in order to suppress the contribution from the surface-attached radon progeny. With these selections, the total data exposure is 100.7 tonne$\cdot$day. There are 2111 final candidate events.
The major backgrounds include: 1) ER events from the tritium, $^{85}$Kr, $^{127}$Xe, $^{136}$Xe, and flat ER~($^{222}$Rn, material radioactive isotopes, solar $\nu$); 2) NR events from neutrons; 3) accidental background. The expected background contribution is $2133.5\pm 74.6$~\cite{Zhou2021}.
The distribution of these events is shown in Fig.~\ref{fig:2D}, overlaid with the signal distribution of inelastic scattering of $m_\chi=1$~TeV and $\delta=200$ keV.
As a reference, within the 68$\%$ contour of this signal distribution, the number of events is $32$ in the data and the expected background is $43.6^{+29.1}_{-15.7}$.

\section{Signal test and results}
\label{fitting}
A test statistics based on the profile likelihood ratio (PLR)~\cite{Junk1999,read2000modified,Read2002,Cowan2011} is constructed, and the unbinned likelihood function takes the same form as that defined in Ref.~\cite{PandaX-II:2017hlx}. We use the standard CL$_{s+b}$ method to test the background-only and background-plus-signal hypotheses on the data. In most of the parameter space of the DM model considered here, the interaction is dominated by either the tree-level inelastic scattering or the  loop-level elastic scattering. Therefore, signals from these two processes are tested individually, to give relatively conservative limits.

\begin{figure}[htbp]
\centering
\includegraphics[width=0.9\linewidth]{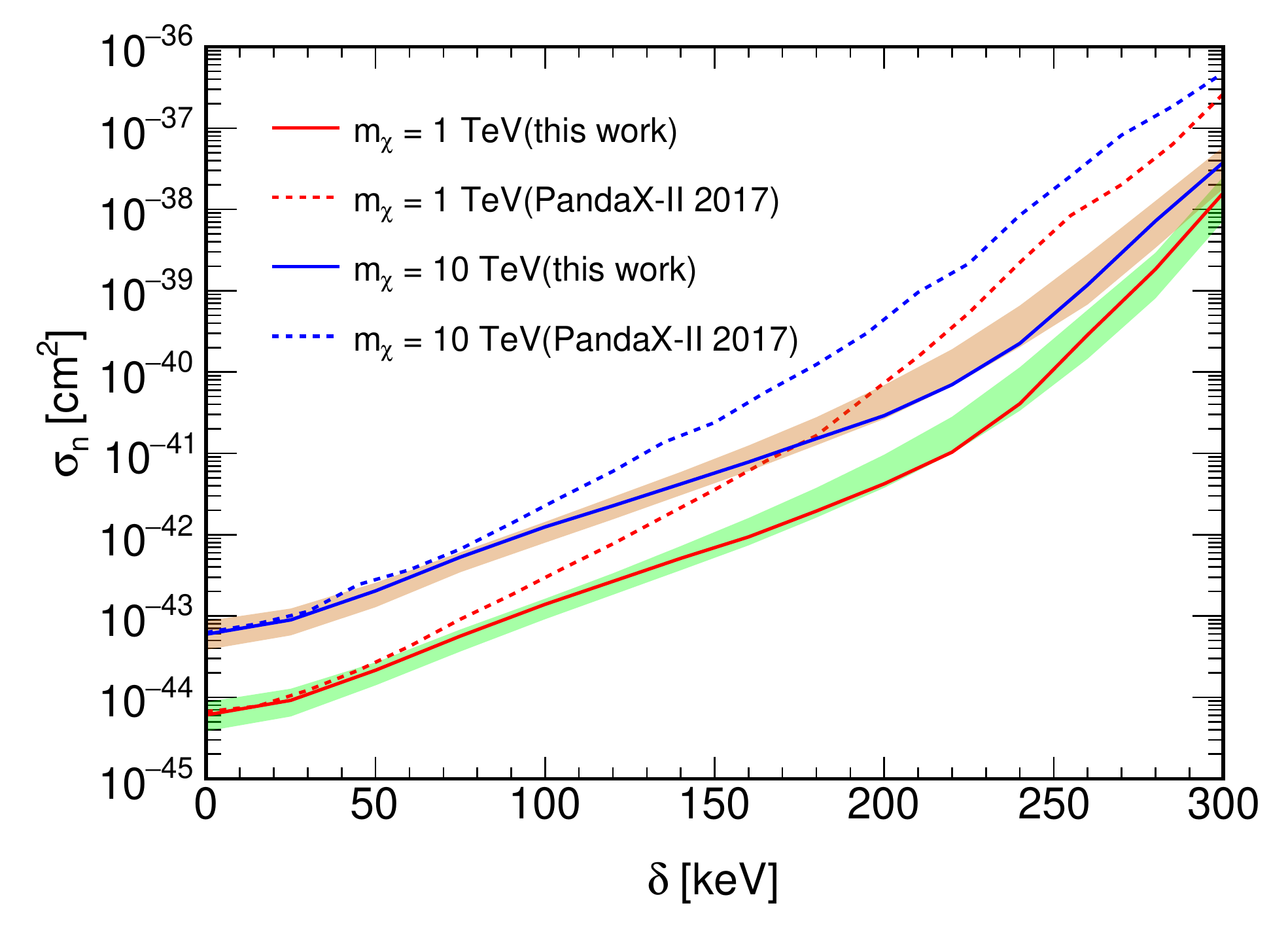}
\caption{The $90\%$ C.L. upper limits of the SI DM-nucleon cross section $\sigma_n$ (solid lines) as a function of $\delta$ at the dark matter masses of 1 (red) and 10~TeV (blue), with the $\pm1\sigma$ sensitivity bands stacked (green for 1 TeV, orange for 10~TeV). The corresponding limits in Ref.~\cite{Chen2017} are plotted as dashed lines with the same colors.}
\label{fig:1Tor10TLimit}
\end{figure}
For the tree-level inelastic scattering, the 90\% confidence level (C.L.) upper limits on the DM-nucleon cross section $\sigma_n$ as a function of the mass splitting $\delta$ for the DM masses of 1~TeV and 10~TeV are obtained. The results are shown in Fig.~\ref{fig:1Tor10TLimit}. 
At $\delta=50$~keV, the sensitivity is improved by a factor of 1.5 and 1.4 for $m_\chi=1$~TeV and $10$~TeV respectively, in comparison with the results in Ref.~\cite{Chen2017}. For $\delta=200$~keV, larger improvements are observed, with a factor of 10.5 for $m_\chi=1$~TeV and 8.7 for $m_\chi=10$~TeV,  partly resulted from the extended searching window.

 \begin{figure}[htbp]
\includegraphics[width=0.9\linewidth]{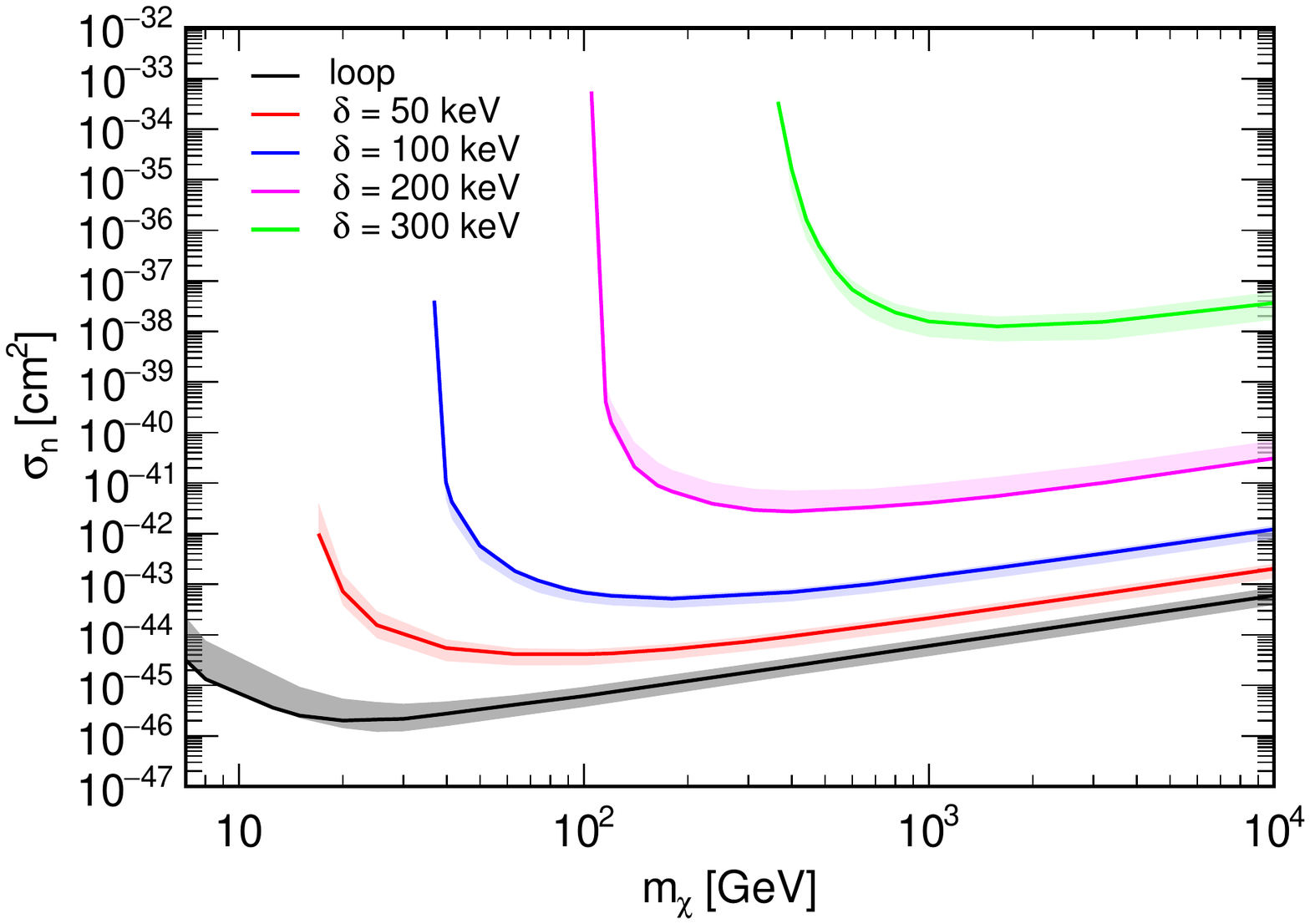}
\caption{The $90\%$ C.L. upper limits of the DM-nucleon cross section $\sigma_n$ as a function of DM mass $m_\chi$ for different mass splitting $\delta$, with the $\pm1\sigma$ sensitivity band. Loop-level elastic scattering results are also presented. }
\label{fig:gap100Limit}
\end{figure}

\begin{figure}[htbp]
\includegraphics[width=0.9\linewidth]{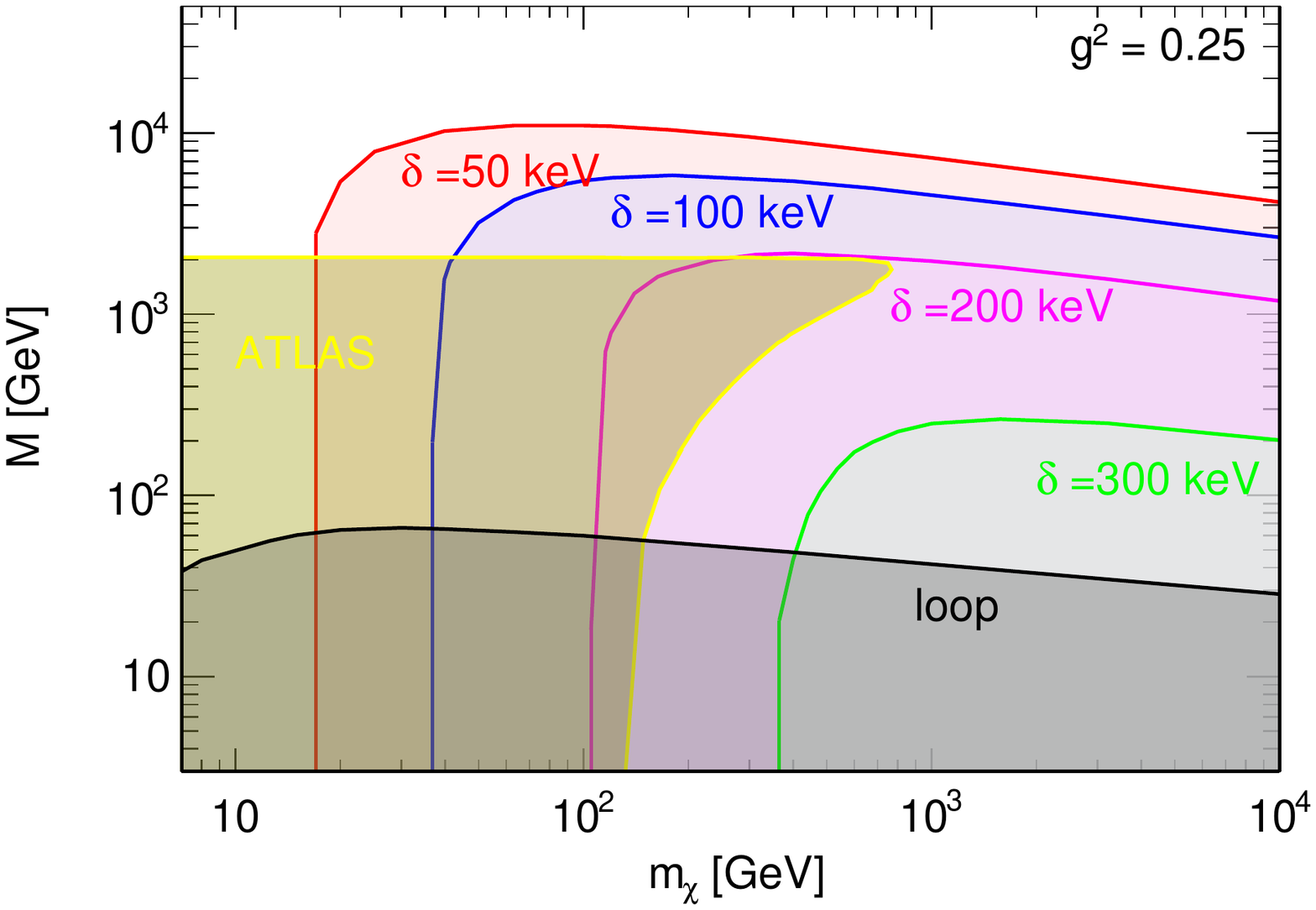}
\caption{Constraints on the mediator mass $M$ extracted from the PandaX-II data, assuming $g^2 = 0.25$. The exclusion parameter phase for $m_\chi$-M from ATLAS is presented in the yellow zone with $g_\chi=1$ and $g_q=0.25$.  }
\label{fig:MvLimit}
\end{figure}

For the loop-level elastic scattering, the corresponding signal spectrum is identical to the standard spin-independent DM-nucleon elastic scattering, the derived upper limits on  $\sigma_n$ from the extended signal region are similar to Ref.~\cite{PandaX-II:2020oim} as the elastic scattering signal is mainly at low energy region.
Strong downward fluctuation is observed for the low DM mass region of $m_\chi<10$~GeV, and the limit is power constrained~\cite{2011Power1} to the $-1\sigma$ of the sensitivity band. 
{\color{black} Due to the reduction FV~\cite{Zhou2021} and the updated light yield and charge yield responses~\cite{PandaX-II:2021jmq}, the exclusion limits at loop-level in this work is more conservative than Ref.~\cite{PandaX-II:2020oim}}. 
The upper limits as a function of DM mass from tree-level processes with fixed mass splitting of $\delta=50$, 100, 200 and 300~keV, are presented in Fig.~\ref{fig:gap100Limit}.
For large $\delta$, limits can only be set for large DM mass region due to the kinematic limit from halo escape velocity.

In the simplified model with a vector boson mediator, direct constraints on the mediator mass $M$ can be derived from the tree-level and loop-level signal searches respectively. In addition, DM searches at collider experiments can also give constraints on this vector boson mediator. 
The 90$\%$ C.L. lower limits of $M$ as functions of the DM mass are given in Fig.~\ref{fig:MvLimit}, with the assumption of $g^2=0.25$, which is chosen to be comparable with existing collider search directly ($g_\chi=1$ and $g_q=0.25$)~\cite{ATLAS:DMsummary}.
For other values of $g$, the limit on the mediator mass $M$ scales accordingly based on Eqs.~\ref{eq:L_tree} and~\ref{eq:L_loop}.
At non-zero mass splitting $\delta$, stringent constraints on the mediator mass $M$ are obtained from the tree-level inelastic process for larger DM mass. The loop level elastic process provides complementary constraints at the small DM mass region, which extends to higher DM mass with the increasing of $\delta$.  
Searches on collider also provide constraints on parameter space, particularly for relatively small DM mass.
The exclusion region from the mono-X ($E_{\rm T}^{miss}+$visible particles) searches with the ATLAS detector at the LHC is also presented in Fig.~\ref{fig:MvLimit} as a complement.
Under the aforementioned coupling constant assumption, for a mass splitting less than 200~keV, the combined direct detection and collider searches have excluded the parameter space for pDDM with a vector mediator mass less than 2 TeV. 
On the other hand, the elastic scattering provides competitive constraints for large DM mass ($>100$~GeV), where the collider constraint is inaccessible and the mass splitting is too large for the inelastic scattering to happen.

\section{Summary}
\label{sec:summary}
In summary, we performed the analysis to search for the two-component Majorana DM signal with the 100.7 ton$\cdot$day full exposure data of PandaX-II. 
Such a model offers a natural suppression to the traditional elastic scattering process as it happens only at the one-loop level.
In the simplified model, both the tree-level inelastic and loop-level elastic processes contribute to the NR signal in direct detection experiment. In comparison with previous PandaX-II work, the signal window is extended to higher NR energy region to enhance the sensitivity to high mass splitting values.  
No significant excess is observed, thus the limits on the DM-nucleon cross section $\sigma_n$ as functions of DM masses and mass splitting values are obtained. 
We also derive the constraints on the mass of the vector boson mediator within the framework of pDDM.
A larger DM mass region is covered with the contribution from the loop-level elastic process. 
In combination with results from collider experiments, more stringent constraints are applied to this theoretical model.
Currently, the next generation detector, PandaX-4T is taking data, which will provide data with larger exposure and energy range to explore this dark matter scenario further.

\input{acknowledgement}

\bibliography{ref}
 \end{document}

%% file: authorlist.tex

\def\shKeyLab{School of Physics and Astronomy, Shanghai Jiao Tong University, MOE Key Laboratory for Particle Astrophysics and Cosmology, Shanghai Key Laboratory for Particle Physics and Cosmology, Shanghai 200240, China}
\def\USTClab{State Key Laboratory of Particle Detection and Electronics, University of Science and Technology of China, Hefei 230026, China}
\def\USTCdep{Department of Modern Physics, University of Science and Technology of China, Hefei 230026, China}
\def\BUAA{School of Physics, Beihang University, Beijing 102206, China}
\def\BUAALab{Beijing Key Laboratory of Advanced Nuclear Materials and Physics, Beihang University, Beijing, 102206, China}
\def\zzu{School of Physics and Microelectronics, Zhengzhou University, Zhengzhou, Henan 450001, China}
\def\pku{School of Physics, Peking University, Beijing 100871, China}
\def\YaLongSD{Yalong River Hydropower Development Company, Ltd., 288 Shuanglin Road, Chengdu 610051, China}
\def\IAP{Shanghai Institute of Applied Physics, Chinese Academy of Sciences, 201800 Shanghai, China}
\def\CHEPpku{Center for High Energy Physics, Peking University, Beijing 100871, China}
\def\SDUdep{Research Center for Particle Science and Technology, Institute of Frontier and Interdisciplinary Science, Shandong University, Qingdao 266237, Shandong, China}
\def\SDUlab{Key Laboratory of Particle Physics and Particle Irradiation of Ministry of Education, Shandong University, Qingdao 266237, Shandong, China}
\def\UMD{Department of Physics, University of Maryland, College Park, Maryland 20742, USA}
\def\TDLee{Tsung-Dao Lee Institute, Shanghai Jiao Tong University, Shanghai, 200240, China}
\def\MESJTU{School of Mechanical Engineering, Shanghai Jiao Tong University, Shanghai 200240, China}
\def\SYU{School of Physics, Sun Yat-Sen University, Guangzhou 510275, China}
\def\NKU{School of Physics, Nankai University, Tianjin 300071, China}
\def\FDU{Key Laboratory of Nuclear Physics and Ion-beam Application (MOE), Institute of Modern Physics, Fudan University, Shanghai 200433, China}
\def\USST{School of Medical Instrument and Food Engineering, University of Shanghai for Science and Technology, Shanghai 200093, China}
\def\SJTUSC{Shanghai Jiao Tong University Sichuan Research Institute, Chengdu 610213, China}
\def\Princeton{Physics Department, Princeton University, Princeton, NJ 08544, USA}
\def\MIT{Department of Physics, Massachusetts Institute of Technology, Cambridge, MA 02139, USA}
\def\SARI{Shanghai Advanced Research Institute, Chinese Academy of Sciences, Shanghai 201210, China}
\def\SPEIT{SJTU Paris Elite Institute of Technology, Shanghai Jiao Tong University, Shanghai, 200240, China}


\author{Ying Yuan}\affiliation{\shKeyLab}
\author{Abdusalam Abdukerim}
\author{Zihao Bo}
\author{Wei Chen}\affiliation{\shKeyLab}
\author{Xun Chen}\email[Corresponding author: ]{chenxun@sjtu.edu.cn}\affiliation{\shKeyLab}\affiliation{\SJTUSC}
\author{Yunhua Chen}\affiliation{\YaLongSD}
\author{Chen Cheng}\affiliation{\SYU}
\author{Xiangyi Cui}\affiliation{\TDLee}
\author{Yingjie Fan}\affiliation{\NKU}
\author{Deqing Fang}
\author{Changbo Fu}\affiliation{\FDU}
\author{Mengting Fu}\affiliation{\pku}
\author{Lisheng Geng}\affiliation{\BUAA}\affiliation{\BUAALab}\affiliation{\zzu}
\author{Karl Giboni}
\author{Linhui Gu}\affiliation{\shKeyLab}
\author{Xuyuan Guo}\affiliation{\YaLongSD}
\author{Ke Han}\affiliation{\shKeyLab}
\author{Changda He}\affiliation{\shKeyLab}
\author{Jinrong He}\affiliation{\YaLongSD}
\author{Di Huang}\affiliation{\shKeyLab}
\author{Yanlin Huang}\affiliation{\USST}
\author{Zhou Huang}\affiliation{\shKeyLab}
\author{Ruquan Hou}\affiliation{\SJTUSC}
\author{Xiangdong Ji}\affiliation{\UMD}
\author{Yonglin Ju}\affiliation{\MESJTU}
\author{Chenxiang Li}\affiliation{\shKeyLab}
\author{Mingchuan Li}\affiliation{\YaLongSD}
\author{Shu Li}\affiliation{\MESJTU}
\author{Shuaijie Li}\affiliation{\TDLee}
\author{Qing  Lin}\affiliation{\USTClab}\affiliation{\USTCdep}
\author{Jianglai Liu}\email[Spokesperson: ]{jianglai.liu@sjtu.edu.cn}\affiliation{\shKeyLab}\affiliation{\TDLee}\affiliation{\SJTUSC}
\author{Xiaoying Lu}\affiliation{\SDUdep}\affiliation{\SDUlab}
\author{Lingyin Luo}\affiliation{\pku}
\author{Wenbo Ma}\affiliation{\shKeyLab}
\author{Yugang Ma}\affiliation{\FDU}
\author{Yajun Mao}\affiliation{\pku}
\author{Yue Meng}\affiliation{\shKeyLab}\affiliation{\SJTUSC}
\author{Nasir Shaheed}\affiliation{\SDUdep}\affiliation{\SDUlab}
\author{Xuyang Ning}\affiliation{\shKeyLab}
\author{Ningchun Qi}\affiliation{\YaLongSD}
\author{Zhicheng Qian}\affiliation{\shKeyLab}
\author{Xiangxiang Ren}\affiliation{\SDUdep}\affiliation{\SDUlab}
\author{Changsong Shang}\affiliation{\YaLongSD}
\author{Guofang Shen}\affiliation{\BUAA}
\author{Lin Si}\affiliation{\shKeyLab}
\author{Wenliang Sun}\affiliation{\YaLongSD}
\author{Andi Tan}\affiliation{\UMD}
\author{Yi Tao}\affiliation{\shKeyLab}\affiliation{\SJTUSC}
\author{Anqing Wang}\affiliation{\SDUdep}\affiliation{\SDUlab}
\author{Meng Wang}\affiliation{\SDUdep}\affiliation{\SDUlab}
\author{Qiuhong Wang}\affiliation{\FDU}
\author{Shaobo Wang}\affiliation{\shKeyLab}\affiliation{\SPEIT}
\author{Siguang Wang}\affiliation{\pku}
\author{Wei Wang}\affiliation{\SYU}
\author{Xiuli Wang}\affiliation{\MESJTU}
\author{Zhou Wang}\affiliation{\shKeyLab}\affiliation{\SJTUSC}\affiliation{\TDLee}
\author{Mengmeng Wu}\affiliation{\SYU}
\author{Weihao Wu}
\author{Jingkai Xia}\affiliation{\shKeyLab}
\author{Mengjiao Xiao}\affiliation{\UMD}
\author{Xiang Xiao}\affiliation{\SYU}
\author{Pengwei Xie}\affiliation{\TDLee}
\author{Binbin Yan}\affiliation{\shKeyLab}
\author{Xiyu Yan}\affiliation{\USST}
\author{Jijun Yang}
\author{Yong Yang}\affiliation{\shKeyLab}
\author{Chunxu Yu}\affiliation{\NKU}
\author{Jumin Yuan}\affiliation{\SDUdep}\affiliation{\SDUlab}
\author{Dan Zhang}\affiliation{\UMD}
\author{Minzhen Zhang}\affiliation{\shKeyLab}
\author{Peng Zhang}\affiliation{\YaLongSD}
\author{Tao Zhang}
\author{Li Zhao}\affiliation{\shKeyLab}
\author{Qibin Zheng}\affiliation{\USST}
\author{Jifang Zhou}\affiliation{\YaLongSD}
\author{Ning Zhou}\email[Corresponding author: ]{nzhou@sjtu.edu.cn}\affiliation{\shKeyLab}
\author{Xiaopeng Zhou}\affiliation{\BUAA}
\author{Yong Zhou}\affiliation{\YaLongSD}

\collaboration{PandaX Collaboration}
\noaffiliation

%% file: acknowledgement.tex

\section{Acknowledgement}
 
We thank Tong Li and Wei Chao for helpful discussions. This project is supported in part by a grant from the Ministry of Science and Technology of
China (No. 2016YFA0400301), grants from National Science
Foundation of China (Nos. 12090060, 12005131, 11905128, 11925502, 11735003, 11775141), 
and by Office of Science and
Technology, Shanghai Municipal Government (grant No. 18JC1410200). We thank supports from Double First Class Plan of
the Shanghai Jiao Tong University. We also thank the sponsorship from the Chinese Academy of Sciences Center for Excellence in Particle
Physics (CCEPP), Hongwen Foundation in Hong Kong, Tencent
Foundation in China and Yangyang Development Fund. Finally, we thank the CJPL administration and
the Yalong River Hydropower Development Company Ltd. for
indispensable logistical support and other help.